\input harvmac

\input epsf
\ifx\epsfbox\UnDeFiNeD\message{(NO epsf.tex, FIGURES WILL BE
IGNORED)}
\def\figin#1{\vskip2in}
\else\message{(FIGURES WILL BE INCLUDED)}\def\figin#1{#1}\fi
\def\ifig#1#2#3{\xdef#1{fig.~\the\figno}
\goodbreak\topinsert\figin{\centerline{#3}}%
\smallskip\centerline{\vbox{\baselineskip12pt
\advance\hsize by -1truein\noindent{\bf Fig.~\the\figno:} #2}}
\bigskip\endinsert\global\advance\figno by1}


\def\a{\alpha}

\def \om {\omega}

\def \del {\partial}

\def\ec{ {\cal E} }
\def \const {{\rm const}}
\def \ha  { {\textstyle{1\ov 2} } }

\def\bs{\bigskip }

\def \s {\sigma}

\def \td {\tilde }

\def \ov {\over }

\def\rr{R_{11}}
\def\gef{ g_{\rm eff} }

\def \lr { \lref}
\def\np {{  Nucl. Phys. }}
\def \pl {{  Phys. Lett. }}

\baselineskip8pt
\Title{
\vbox
{\baselineskip 6pt{\hbox{ }}
{\hbox{hep-th/0101132}} {\hbox{
  }}} }
{\vbox{\centerline {Free Energy and Critical Temperature}
\vskip4pt
 \centerline {in Eleven Dimensions}
}}
\vskip -27 true pt
\centerline  {  Jorge G. Russo{\footnote {$^*$} {e-mail address:
russo@df.uba.ar
 } } 
}

\bigskip

 \centerline {\it  Departamento de F\'\i sica, Universidad de Buenos Aires, }
\smallskip
\centerline {\it  Ciudad Universitaria, Pab. I, 1428 Buenos Aires.}

\medskip\bigskip

\centerline {\bf Abstract}
\medskip
\baselineskip10pt
\noindent

We compute the one-loop contribution to the free energy in eleven-dimensional
supergravity, with the  eleventh dimension compactified on a circle of radius $\rr $.
We find a finite result, which, in a small radius expansion, 
has the form of the type IIA supergravity free energy
plus non-perturbative corrections in the string coupling $g_A$, whose coefficients we determine. 
We then study type IIA superstring theory at finite temperature in the strong coupling
regime by considering M-theory on $R^9\times T^2$, one of the sides of 
the torus being the euclidean time direction, where fermions obey antiperiodic
boundary conditions. 
We find that a certain winding membrane state becomes tachyonic above some  critical temperature, which depends on $g_A$.
At weak coupling, it coincides with the Hagedorn temperature, at large coupling it becomes
$T_{\rm cr} \cong  0.31\ l_P^ {-1} $ (so it is very small in string units).

\medskip
\Date {January 2001}
\noblackbox
\baselineskip 14pt plus 2pt minus 2pt

\lr\aw {J.J.~Atick and E. Witten, \np {\bf B310} (1988) 291.}

\lr\hagedorn{R. Hagedorn, Nuovo Cimento Suppl. {\bf 3} (1965) 147.}

\lr \ggv { M.B. Green, M.  Gutperle and P.  Vanhove, \pl {\bf B409}
(1997) 177, hep-th/9706175.}

\lr\rutse{J.G. Russo and A.A. Tseytlin, \np  {\bf B508} (1997) 245, hep-th/9707134.}

\lr\gubser{S.~S.~Gubser, S.~Gukov, I.~R.~Klebanov, M.~Rangamani and E.~Witten, ``The Hagedorn transition in non-commutative open string theory", hep-th/0009140.}

\lr\fratse{E.~S.~Fradkin and A.~A.~Tseytlin,
Nucl.\ Phys.\ {\bf B227} (1983) 252.}

\lr\sathia{I. Kogan, {\it JETP Lett.} {\bf 45} (1987) 709;
B.~Sathiapalan, Phys. Rev. {\bf D35} (1987) 3277.}

\lr\rohm{R. Rohm, \np {\bf B237} (1984) 553.}

\lr\berg{O. Bergman and M.R. Gaberdiel,  JHEP {\bf 9907} (1999) 022, hep-th/9906055.}

\lr\horava{M.~Fabinger and P.~Horava, 
\np {\bf B580} (2000) 243, hep-th/0002073.}

\lr\bergsh{E. Bergshoeff,  E. Sezgin and P.K. Townsend, 
Ann. Phys. {\bf 185} (1988) 330.}

\lr\hardix{L.J.~Dixon and J.A.~Harvey, \np {\bf B264} (1986) 93;
N.~Seiberg and E.~Witten, \np {\bf B276} (1986) 272. }



\lr\giddings{M. Bowick and S. Giddings, Nucl.~Phys. {\bf B325} (1989) 631.}


\lr\amati{D. Amati and J.G. Russo, Phys.~Lett. {\bf B454} (1999) 207, hep-th/9901092.}

\lr\bakas{I.~Bakas, A.~Bilal, J.~P.~Derendinger and K.~Sfetsos,
Nucl.\ Phys.\ B {\bf 593}, 31 (2001),
hep-th/0006222 .}

\lr\chau{S. Chaudhuri,  ``Deconfinement and the Hagedorn transition in string theory",
hep-th/0008131.}

\lr\odin{
A.A. Bytsenko and S.A. Ktitorov, Phys.~Lett. {\bf B225} (1989) 325;
A.A. Bytsenko and S.D. Odintsov, Phys.~Lett. {\bf B243} (1990) 63.}


\lr\dewitt{B. de Wit, J. Hoppe and H. Nicolai, 
\np  {\bf B305} [FS 23] (1988) 545.}

\lr\banks{
T.~Banks, W.~Fischler, S.~H.~Shenker and L.~Susskind,
Phys.\ Rev.\  {\bf D55}, 5112 (1997),
hep-th/9610043.}

\lr\liu{
H.~Liu and A.~A.~Tseytlin,
JHEP {\bf 9801}, 010 (1998),
hep-th/9712063.}

\lr\meana{ 
M.~L.~Meana, M.~A.~Osorio and J.~P.~Pe\~ nalba,
Nucl. Phys.  {\bf B531}, 613 (1998),
hep-th/9803058.}

\lr\chm{S.~Chaudhuri and D.~Minic,
Phys.\ Lett.\  {\bf B433}, 301 (1998),
hep-th/9803120.}

\lr\sath{
B.~Sathiapalan,
Mod.\ Phys.\ Lett.\ A {\bf 13}, 2085 (1998),
hep-th/9805126.}


\lr\barbon{J.~L.~Barbon, I.~I.~Kogan and E.~Rabinovici,
Nucl.\ Phys.\ B {\bf 544}, 104 (1999),
hep-th/9809033.}

\lr\rama{
S.~Kalyana Rama and B.~Sathiapalan,
Mod.\ Phys.\ Lett.\ A {\bf 13}, 3137 (1998),
hep-th/9810069.}

\lr\bst{E. Bergshoeff,  E. Sezgin and Y. Tanii, \np {\bf B298} (1988) 187.}

\lr\schw{J.H. Schwarz, \pl {\bf B360} (1995) 13,
hep-th/9508143.}

\lr\duf{M.J. Duff, T. Inami, C.N. Pope, E. Sezgin and K.S. Stelle, 
\np {\bf B227} (1988) 515.}

\lr\russo{J.G. Russo, in: Proceedings of the APCTP Winter School ``Dualities in Gauge and String Theories" (Eds. Y.M. Cho and S. Nam) World Scientific (1997),
hep-th/9703118.}

\lr\constru{J.G. Russo, \np {\bf B535} (1998) 116, hep-th/9802090.}

\lr\fuji{K. Fujikawa and J. Kubo, Phys. Lett. {\bf B199} (1987) 75.}

\lr\terras {A. Terras, {\it Harmonic Analysis on Symmetric Spaces and Applications}, vol. I, Springer-Verlag (1985).}

\lr\dln{B. de Wit, M. L\" uscher and H.~Nicolai, \np {\bf B320} (1989) 135.}

\lr\plefka {B.~de Wit, K.~Peeters and J.~Plefka, Phys.~Lett.~{\bf B409} (1997) 117.}


\lr \green{M.B. Green and M. Gutperle, hep-th/9604091.}
\lr\mgreen{M.B. Green, hep-th/9712195.}

 \lr\rusty{J.G. Russo and A.A. Tseytlin, \np B490 (1997) 121.}

\lr \bergsho{ E. Bergshoeff, E.  Sezgin and P.K. Townsend, \pl B189 (1987)
75.}


\lr\trusso{J.G. Russo, \pl B400 (1997) 37, hep-th/9701188.}

\lr \gsb {
 M.B. Green, J.H. Schwarz and L. Brink,  \np B198 (1982) 474. }

\lr \gsh {M.B. Green  and J.H. Schwarz, \np B198 (1982) 441.}

 \lr \gv {M.B. Green and P. Vanhove, \pl B408 (1997) 122,
hep-th/9704145.}

\lr \grgu{M.B. Green and M. Gutperle, 
\np B498 (1997) 195, hep-th/9701093.}

\lr\duff{M.J. Duff, P.S. Howe, T. Inami and K.S. Stelle, 
\pl B191 (1987) 70.}

\newsec{Introduction}

A problem of interest in type IIA superstring theory is to understand the evolution of the degrees of freedom of the system as the coupling is increased from weak to strong values.
For weak couplings, the theory can be described in terms of a supersymmetric
relativistic string, but for strong coupling the relevant degrees of freedom are not well
understood.
The study of a system at finite temperature can give some non-trivial information
about its microscopic degrees of freedom, and about their behavior as the system is heated up to high temperatures.
In this paper we will discuss some features of M-theory at finite temperature,
with the eleventh dimension $X^{11}$ compactified on a circle of radius $R_{11}$.
Given our limited knowledge of M-theory, a complete treatment is of course presently impossible.
Nevertheless, we will find, somewhat surprisingly, 
 that very interesting aspects can be revealed by simple calculations.

For small radius $R_{11}$, one must recover the thermodynamics of string theory, which
is well understood qualitatively \aw . There is a first order  phase transition at some 
$T_0<T_{H}$, where $T_H$ is the Hagedorn temperature \hagedorn , with a large latent heat
leading to a gravitational instability of the thermal ensemble.
The way the Hagedorn temperature is calculated in string theory basically involves free string theory methods, so an
important question is how interactions  modify this picture. In particular, one would like to understand what happens to the Hagedorn temperature in the type IIA 
superstring theory  at strong coupling, in other words, what is the fate of the Hagedorn transition in $D=11$.

The presence of phase transitions is usually reflected
as infrared divergences in the one-loop free energy. 
This calculation is difficult to generalize to eleven dimensions, for a number of reasons. 
In string theory,
the one-loop free energy is essentially the sum over free energy contributions of each individual physical string mode.
In order to compute a one-loop free energy in M-theory in this way, a more detailed knowledge of the relevant physical degrees of freedom would be required.
In addition, 
in the eleven dimensional theory there is no
coupling constant parameter, and higher loops 
will give contributions to the free energy of order one.
Nevertheless,  as a first step, one could  try to determine the free energy in eleven-dimensional supergravity. 
As a physical application, 
the eleven-dimensional supergravity result may be then used to incorporate  non-perturbative D0 brane contributions to the one-loop free energy of string theory.
$D=11$ supergravity is not renormalizable as a quantum field theory. However, as we show
in section 2, the one-loop free energy is  finite, thanks to a cancellation between boson and fermion contributions (a general discussion of one-loop divergences in
supergravity  can be found in \fratse ).
This is important, since the presence of an UV divergence (as happens in a purely bosonic theory), would prevent us from recovering
the ten-dimensional physics in the small radius limit.

One of the difficulties in understanding details of the Hagedorn transition in string theory is that gravitational effects cannot be neglected shortly above the Hagedorn temperature, due to a large genus zero contribution to the free energy \aw . The description in terms of a free string gas in a flat background is not applicable; rather,  one expects gravitational collapse near the Hagedorn temperature. 
One can give a qualitative  description
of what physical picture should be expected (e.g. by using the microcanonical ensemble one can argue which string configurations are favored and dominate the density of states), but the arguments are mostly based on string propagation in flat space. A more detailed understanding was recently achieved in non-commutative open string theory, which does not contain gravitation \gubser .

In the case of M theory at large radius $R_{11}$, a flat-theory description of the thermal ensemble
can be justified only for temperatures $T\ll O(l_P^{-1})$, where
$l_P$ is the eleven-dimensional Planck length.
The reason is the following one. Statistical mechanics is valid provided the system has
a large volume, so that it contains many degrees of freedom.
In eleven dimensions, large volume means that the size $R$ of the system is
$R\gg l_P$. On the other hand, a flat theory description requires that corrections to the flat Minkowski metric are small, i.e. ${G_NE\over R^8}\ll 1$, with $G_N\sim l_P^9$. Now consider the thermal ensemble at temperature $T=O(l_P^{-1})$. Then the energy density is ${E\over R^{10}}=\const. l_P^{-11} $, since there is no other parameter in the theory. Hence gravitational effects are of order
${G_NE\over R^8}\sim { R^2\over l_P^2}$, so  they are important for $R>l_P$. Thus it is not possible to have a statistical description of the thermal ensemble in flat space
near the Planck temperature. A temperature $T=O(l_P^{-1})$ is in fact the maximum temperature that a statistical system can reach in eleven dimensions without gravitational collapse. For any $T>O(l_P^{-1})$,
a thermal ensemble with size $R>l_P$ will be inside its Schwarzschild radius, so it will collapse into a black hole.
This can be compared with the situation
in string theory, where one has the string coupling $g_s$ as a free parameter, and for sufficiently small $g_s$  gravity can be ignored at any $T<T_H$
(but not at $T>T_H$ due to the appearance of a genus zero 
contribution $F_0\sim -1/g_s^2$ \aw ).

Despite these complications, using flat-space methods to study the
Hagedorn transition has led to important insights on the nature of string theory and its physical degrees of freedom.  
One may then expect that a similar simplified study in eleven dimensions can teach us important lessons about M-theory. 
Here we will find that the Hagedorn temperature admits a straightforward generalization 
to eleven dimensions. In string theory, the
 Hagedorn temperature  can be found as the temperature at which a certain winding state becomes tachyonic \sathia \ (see sect.~3). In eleven dimensions, this 
winding state is a winding membrane. 
We find that it becomes tachyonic at some critical temperature $T_{\rm cr}=T_{\rm cr}(g_A)$, where $g_A$ is the type IIA string coupling, which smoothly interpolates between the Hagedorn temperature ($g_A\ll 1$)  and a critical temperature $T=O(l_P^{-1})$ ($g_A\gg 1$). This is done in sect.~4, where we also include some remarks about  a duality to type 0A string theory~\hardix . 

Other discussions about the Hagedorn temperature in string theory can be found e.g. in \refs{\giddings\amati\bakas - \chau }. 
There have also been some discussions on membrane theory
at finite temperature in  \odin , where there is an attempt of computing the free energy. 
The matrix theory approach \banks\ has also been considered at finite temperature in 
refs.~\refs{\liu\meana\chm -\sath }, which  discuss other aspects of the Hagedorn transition.
There is no overlap with the present work. 


In appendix A we review standard properties of the non-holomorphic Eisenstein 
series~\terras . Appendix B contains new Eisenstein-type series with alternating signs in the sum. We derive 
formulas for the expansions at large and small values of the modular parameter. In the main text, these series arise as the contribution of antiperiodic fermions.

\newsec{Free energy in eleven dimensions}

The thermal ensemble at temperature $T$ can be studied
as usual by considering the theory in Euclidean space where the
time coordinate is compactified on a circle of circumference  $1/T$, i.e.
\eqn\uno{
X^0=X^0+2\pi R_0\ ,\ \ \ \ \ T=(2\pi R_0)^{-1}\ ,
}
where fermions obey antiperiodic boundary conditions. In order
to have a description of type IIA superstring theory which can be extended 
beyond perturbation theory, 
one should thus  consider  euclidean M-theory on
a 2-torus $X^0, X^{ 11} $, 
$$
X^ 0= X^0 + 2 \pi R_0 \ ,\ \ \ \ \ X^ { 11} = X^ { 11} + 2\pi \rr \ .
$$
where fermions are antiperiodic around $X^0$, and periodic  around $X^{11}$. 
Here we will consider the case of a rectangular torus.


The supersymmetric compactification of M-theory on a 2-torus gives rise to a theory that inherits the $SL(2,Z)$
isometry of the 2-torus, because this symmetry is not broken by boundary conditions of the
fields. In the case of finite temperature M-theory,
the different boundary conditions for fermions in the directions $X^0,X^{11}$ break the 
$SL(2,Z)$ symmetry.
 This will be reflected in the present calculation of the free energy.
Instead, in a purely bosonic theory, the exact partition function must have the symmetry 
\eqn\juno{
Z_{ \rm bos }(\gef )=Z_{ \rm bos} (1/\gef )\ ,\ \ 
}
with
\eqn\ccc{
 \gef \equiv  {R_{11}\over R_0}=2\pi \sqrt{\a '} T g_A\ .
}
Here  $g_A=\rr/\sqrt {\a' } $ is the type IIA string coupling.
This implies for the free energy the relation:
\eqn\yuno {
F_ { \rm bos} (\gef ,A ,l_P)= \gef F_{ \rm bos} ({1\over\gef },A,l_P)
\ ,
}
where $A=R_{0}R_{11}$. 
The symmetry \yuno\ can also be expressed in  terms of string 
theory parameters $\{ g_A,T,\a' \} $ and relates low and high 
temperature regimes, as well as weak and strong coupling regimes.

Here we will obtain the one-loop contribution to the free energy in eleven-dimensional  supergravity compactified on a circle
by adding to the ten-dimensional supergravity expression  an extra factor
containing a sum over Kaluza-Klein modes $\sum_m e^{-\pi \tau m^2/R^2}$.
The free energy in ten-dimensional supergravity  can be obtained from  superstring
theory 
as a limit $\a' \to 0$ (deriving the one-loop contribution to the free energy directly from
the component formulation of $D=11$ supergravity is more complicated).
This is similar to \refs{\ggv ,\rutse }, where the one-loop 4-graviton amplitude in $D=11$ supergravity was computed by adding the Kaluza-Klein modes to the $D=10$ supergravity amplitude.

\subsec{Free energy in a simplified model}

Before considering the  free energy in string theory, it is instructive to study a simplified model
in which we sum up the individual free energies of Kaluza-Klein scalar fields. That is
\eqn\frrr{
F(T)=VT \sum_m \int {d^{D-2}p\over (2\pi )^{D-2}} \log\big[1-e^{-\omega_p/T}\big]\ ,\ \ \ \ 
\omega_p^2=\vec p^2+ {m^2\over R_{11}^2} \ .
}
Expanding the logarithm and using 
$$
e^{-2\sqrt{ab} } = \sqrt{b \over \pi} \int_0^\infty {dt\over t^{3/2}} e^{-a t-b/t} \ ,
$$
one can write 
\eqn\rrrr{
F(T)=- V \sum_m \int_0^\infty {dt\over t^{D+1\over 2}}
\sum_{w=1}^\infty 
\exp \big[-{\pi m^2t\over R_{11}^2} -{\pi w^2 R_0^2\over  t}\big]\ ,
\ \ \ \ R_0=(2\pi T)^{-1}\ ,
}
where we have ignored a multiplicative numerical constant.
Let us now set $D=11$ (so that $[V]$=cm$^9$). Including the vacuum part $w=0$, and making a Poisson resummation in $w$, i.e.  
\eqn\pois{
\sum_{w=-\infty }^\infty e^{- {\pi R^2 \over t} w^2  }={\sqrt{t }\over R }
\sum_{k=-\infty }^\infty e^{- \pi t {k^2\over R^2} }\ ,
} 
we get 
\eqn\ccrr{
F(T)=- \pi T V \int_0^\infty {dt \over t ^{11/2} } \sum_{k,m}
\exp \bigg[ -\pi t \big( {k^2\over R_0^2}+ {m^2\over R_{11}^2} \big)\bigg]\ .
}
This can be recognized as the one loop contribution of Kaluza-Klein scalars
associated with the 2-torus $X^0,X^{11}$.
The integral diverges in the ultraviolet region $t\to 0$.
We can isolate the divergent piece by  performing Poisson resummation in $k,m$ and introducing the new integration variable 
$s=1/t $, so that
\eqn\aarr{
F(T)=- \ha VR_{11} \int_0^\infty {ds \over s} s ^{11/2}  \sum_{w,w'}
\exp \bigg[ -\pi s \big( w^2 R_0^2+ {w'}^2 R_{11}^2 \big)\bigg]
}
Now the UV divergence is in the term $(w,w')=0$. In string theory the analog term will
cancel against a fermion contribution. 
Thus we get
\eqn\ain{
F(T)=-  {VR_{11} \Gamma(11/2)\over 2\pi^{11/2}}  \sum_{(w,w')\neq (0,0)} \big( w^2 R_0^2+ {w'}^2 R_{11}^2
\big)^{-{11\over 2}} + \ {\rm divergent\ term}\ .
}
This can be expressed in terms of an Eisenstein series (see appendix A)
\eqn\jjy{
F(T)= -{VR_{11}\Gamma(11/2) \over (\pi R_{11}R_0)^{11\over 2}} \zeta (11) E_{11\over 2}(\gef )+ \ {\rm divergent\ term}\ .
}
It satisfies the symmetry relation mentioned above, $F(\gef ,A )=\gef F(1/\gef ,A)$
(this symmetry still holds for the regularized divergent part for a cutoff 
proportional to $l_P^{-1}$, i.e. independent of the radii $R_0,R_{11}$).
To study the behavior at $\gef\gg 1 $ and $\gef \ll 1$ we use the expansions (A.7), (A.8).  We obtain
\eqn\bosf{
{F(T)\over V}=- {945\zeta (11)\over 32 \pi ^5 R_{11}^{10} }-
{24\zeta(10) \over \pi ^5 R_0^{10}} +O\big( \exp[-2\pi {R_0\over R_{11}} ] \big) \ , \ \ \ \ \ \  R_{11}\ll R_0\ ,
}
\eqn\bosw{
{F(T)\over V} =   -{945\zeta (11) R_{11}\over 32\pi^5 R_0^{11} } -
{24\zeta(10) \over \pi ^5 R_0 R_{11}^9} +O\big( \exp [-2\pi {R_{11}\over R_0} ] \big)  \ , \ \ \ \ \ \ R_{11}\gg R_0\ .
}
The leading term in eq.~\bosw\ has the correct form for the free energy of a massless field theory in $D=11$.
The expression \bosf\ contains two terms with power-like dependence on $R_{11}/R_0$. 
The subleading term proportional to $1/R_0^{10}$ gives the expected  expression for the free energy of
a $D=10$ massless field theory. However, there is a leading term 
 proportional to $1/ \rr ^ {10} $.
 The presence of a term of the form 
$1/ \rr ^ {10} $ in the supergravity calculation at $R_{11}\ll R_0$ would be problematic because
there is no such contribution in superstring theory at weak coupling.
As we shall see below, in the supergravity calculation the analog term  cancels out.

\subsec{One-loop free energy in $D=11$ supergravity}

Let us now consider the supergravity computation. The  calculation of the free energy in type II superstring theory  was carried out in
 \aw \ in the genus one approximation (valid for $g_A\ll 1$), with the result
$$
F_{\rm string}=-{1\over 4}V(4\pi^2\a' )^{-5} \int_{\cal F} {d^2\tau\over \tau_2^6} \ 
\big| \eta (\tau ) \big|^ {-24}
\sum_{w',w}
 e^{-{\pi r_0^2\over\tau_2 } |w'+w\tau |^2}
$$
$$
\times \ \bigg[ ( |\theta_2 |^8   +|\theta_3 |^8+|\theta_4 |^8)(0,\tau )
+ e^{i\pi (w+w')}(  \theta_2 ^4 \bar \theta_4^4  + \theta_4 ^4 \bar \theta_2^4  )(0,\tau ) 
$$
\eqn\ggww{
-\ e^{i\pi w'}(  \theta_2 ^4 \bar \theta_3^4  + \theta_3 ^4 \bar \theta_2^4  )(0,\tau ) 
-e^{i\pi w}(  \theta_3 ^4 \bar \theta_4^4  + \theta_4 ^4 \bar \theta_3^4  )(0,\tau ) 
 \bigg] \ ,\ \ \ \ \ r_0^2\equiv {R_0^2\over\a'}\ .
}
In order to obtain the ten-dimensional supergravity result, we first separate
the term with vanishing winding $w=0$, writing
$$
F_{\rm string}=F_{\rm string}' + F_0
$$
where $F_{\rm string}'$ is as in \ggww\ with the omission of the $w=0$ term in the sum, 
which we call $F_0$. The free energy in ten-dimensional supergravity is obtained by
taking the limit $\a' \to 0$ in $F_0$.
Taking $\a' \to 0$ implies
keeping the leading terms of the theta functions and Dedekind $\eta $ function at large $\tau_2$, that is
$$
\eta(\tau )\cong q^{1/12} (1-q^2)\ ,\ \ \ \ 
\theta_2(0,\tau)\cong 2 q^{1/4}(1+q^2)\ ,\ \ \ \ q\equiv e^{i\pi\tau }\ ,
$$
$$ 
 \theta_3(0,\tau)\cong 1+2 q +O(q^3)\ ,\ \ \ \ 
\theta_4(0,\tau)\cong 1-2q+O(q^3)\ . 
$$
We obtain
\eqn\fra{
F_{\rm SG}^{(10)}=\lim _{\a'\to 0} F_0= - 256 V(4\pi^2\a' )^{-5} \int 
{d\tau_2 d\tau_1\over \tau_2^6} \ \sum_{w'}
\big[1-(-1)^{w'} \big] \ e^{-{\pi r_0^2\over\tau_2 } {w'}^2}\ ,\ \ 
}
where the integration region is now the whole strip $\tau_2>0$, $|\tau_1|<1/2$.

Integrating over $\tau_1$ and making a  Poisson resummation in $w'$, we get
\eqn\pfq{
F_{\rm SG}^{(10)}=-256 {V\over r_0}(4\pi^2\a' )^{-5}
 \int _0^\infty {d\tau_2  \over \tau_2^{11/2}} \ \sum_{k} \big[
 e^{-{\pi \tau_2\over r_0^2} k^2 }- e^{-{\pi \tau_2\over r_0^2}
 (k+{1\over 2})^2 } \big] \ .
}
The appearance of half-integer momentum modes is a well-known distinctive feature of having fermions with antiperiodic boundary conditions. Similar expressions for 
the partition function or for the free energy appear in string compactifications where
fermions obey antiperiodic boundary conditions around some spatial dimension 
\rohm\ (for recent discussions on string compactifications with antiperiodic fermions, see e.g. \refs{\berg , \horava }).
 Taking a similar limit on $F'_{\rm string}$, and making a Poisson resummation in $w'$, gives
\eqn\pff{
{F'_{\rm string}\over VT}\to 
- (4\pi^2\a' )^{-{9\over 2}} \int {d\tau_2  d\tau_1\over \tau_2^{11/2}} \ e^{2\pi\tau_2}
\sum_{k,w}
\big[1-(-1)^w \big] \ e^{-\pi \tau_2 (w^2r_0^2+{k^2\over r_0^2})}\ e^{2\pi i\tau_1 kw} .
}
One can see the presence of the thermal tachyon corresponding to the term $k=0$, $w=\pm 1$,
which reflects as an infrared divergence of $F_{\rm string}$ for $T>T_H$: the integral is
divergent at $\tau_2\to \infty $ for   $r_0^2< 2$ (this is precisely the critical
radius that one obtains by examining the spectrum, see eq.~(3.4) with $a_L=a_R=1/2$). 

The Kaluza-Klein modes associated with  the eleventh dimension have masses $|m|/g_A$ in string units, $m$=integer.
By adding their contribution to $F_{\rm SG}^{(10)} $, given in \pfq\ , we get 
the one-loop contribution
 to the free energy in eleven-dimensional 
supergravity compactified on a circle $X^{11}$. This is
 \eqn\peef{
F_{\rm SG} ^{(11)} =-256 {V\over r_0}  (4\pi^2\a' )^{-5}
 \int _0^\infty {d\tau_2  \over \tau_2^{11/2}} \ \sum_{k,m} 
e^{-{\pi\tau_2\over g_A^2} m^2 } \big[
 e^{-{\pi \tau_2\over r_0^2} k^2 }- 
e^{-{\pi \tau_2\over r_0^2} (k+{1\over 2})^2 } \big] \ .
}
The first term  (containing  $e^{-{\pi \tau_2\over r_0^2} k^2 }$)
is essentially the same as the expression obtained in the previous bosonic example \ccrr . The second term (with 
$e^{-{\pi \tau_2\over r_0^2} (k+{1\over 2})^2 }$) represents the fermion
contribution.
Making Poisson resummation in both $k,m$ we get
 \eqn\peel{
F_{\rm SG}^{(11)} =-256 Vg_A (4\pi^2\a' )^{-5}
 \int _0^\infty {ds  \over s} s^{11/2} \ \sum_{w', n } 
\big[ 1-(-1)^{w'}\big] e^{-\pi s({w'}^2 r_0^2+n^2 g_A^2)}
\ ,
}
i.e. 
\eqn\mas{
F_{\rm SG}^{(11)}  =-256 Vg_A (4\pi^2\a' )^{-5}  {\Gamma(11/2)\over (\pi r_0g_A)^{11/2} }
 \ \sum_{w', n } 
\big[ 1-(-1)^{w'}\big] {\gef ^{11/2}\over ({w'}^2 +n^2 \gef ^2)^{11/2} }
\ ,
}
with $\gef ={R_{11}\over R_0}={g_A\over r_0}$. This can be written in terms of the Eisenstein-type series defined in appendices $A,B$
as follows:
\eqn\mas{
{F_{\rm SG}^{(11)}  \over V }=-  T^{10} {2^9\Gamma(11/2)\zeta(11) \over \pi^{11/2}\gef ^{9/2}}
\big[ E_{11\over 2}(\gef ) -  F_{11\over 2}(\gef ) \big] \ .
}
Using the formulas (A.7), (A.8), (B.11), (B.12) for the weak and strong coupling expansions, we obtain
\eqn\debb{
{F_{\rm SG}^{(11)}  \over V T}= - {24\zeta (10)\over \pi ^5}(2^{10}-1)  \ T^9
\  +  \ O( e^{-2\pi/\gef} )\  ,\ \ \ \gef\ll 1\ ,
}
and
\eqn\fur{
{F_{\rm SG}^{(11)}  \over 2\pi R_{11}V T}=-{945\zeta(11)\over 64\pi^5}(2^{11}-1)\ T^{10}-
{3\ 2^{12}\zeta(10)\over  \pi^{5}}\ {T^{10}\over \gef ^{10}}
 +  O( e^{-2\pi\gef} )\  ,\ \ \  \gef\gg 1 \ .
}
The weak coupling expression \debb\ has the expected field theory behavior ${F\over VT}\sim T^{D-1}$ for a free energy of a $D=10$ dimensional massless field theory. The leading term is in fact  $F_{\rm SG}^{(10)}$ given in \pfq . 
The strong $\gef\gg 1$ coupling expression \fur\ has the expected field theory behavior ${F\over R_{11}VT}\sim T^{D-1}$ for a free energy of a $D=11$ dimensional massless field theory.
This agrees with the expectation that varying $g_A$ from small to large values should lead to an interpolation of a ten-dimensional and an eleven-dimensional theory. This does not happen in the bosonic theory, which
has an extra term at small coupling (the underlying reason for which in the bosonic theory the small radius limit does not give the ten dimensional
result  is  the UV divergence, which is different in ten and eleven dimensions; consequently, some memory of the KK modes survives even at small radius $R_{11}$). 
Note also that the only power-like correction in eq.~\fur\ is  always subleading, since  $\gef >1 $. It is independent of the temperature. Indeed, as a function of $T,g_A$,
the free energy in \fur\ has the form
\eqn\furr{
{F_{\rm SG}^{(11)} \over 2\pi R_{11}V T}=-{945\zeta(11)\over 64\pi^5}(2^{11}-1)\ T^{10}-
{12\zeta(10)\over \pi^{15}} {1\over (\sqrt{\a '}g_A)^{10}}
 +  O( e^{-2\pi\gef } )\  ,\ \   \gef\gg 1  .
}

Finally, one can use the above results to define an improved expression for the free energy of
type IIA superstring theory by adding to the one-loop expression \ggww\ (representing the contribution of perturbative string modes) the contribution of D0 branes
represented by the exponentially small terms in \debb . Their explicit form is obtained using eqs. (A.7), (B.11).
We find
\eqn\fftt{
F_{\rm str+D0}=F_{\rm string} -\ {VT^{10}2^{11} \over \gef  ^5 }\sum_{w,m=1}^\infty \big[ 1-(-1)^w \big] \big( {m\over  w} \big)^5
K_5(2\pi {wm \over\gef} )\ . 
}

\newsec{Hagedorn temperature in string theory}

It is useful to recall the way the Hagedorn temperature in superstring theory is manifested in the spectrum, as the temperature at which a certain winding mode becomes massless \sathia .
We consider the theory in Euclidean space where the
time coordinate $X^0$ is compactified on a circle of circumference  $1/T$. In string theory, the presence of coordinates  compactified on circles 
gives rise to winding string states. The string coordinate $X^0(\sigma,\tau )$
can be expanded as follows:
\eqn\dos{
X^0(\sigma ,\tau )= x^0+ 2\a' p^0 \tau +2 R_0w_0\sigma + \tilde X(\sigma,\tau )\ ,
}
$$
p^0={m_0\over R_0}\ ,\ \ \ \ \ m_0,w_0=0,\pm 1, \pm 2,...
$$
where $\tilde X (\s ,\tau )$ is a single-valued function of $\sigma $ and 
$\int_0^\pi d\s \del_\tau \tilde X^0 =0$.
The hamiltonian and level matching constraints are
\eqn\massa{
H=\a' p_i^2 +{w_0^2R_0^2\over \a' }+\a ' {m^2_0\over R^2_0}
+2(N_L+N_R-a_L-a_R)=0\ ,
}
$$
N_L-N_R=m_0w_0\ .
$$
Here $a_L, a_R$ are the normal ordering constants, which represent the vacuum energy of the 1+1 dimensional field theory
(e.g. for the
bosonic string, $a_L=a_R=1$). 
The Hagedorn temperature can be obtained as usual
by determining the radius $R_0$ at which infrared instabilities first appear.
We have seen this effect in section 2  in the one-loop contribution to the free energy; in the presence of
infrared instabilities, the integral over the torus modular parameter $\tau_2$ diverges at large $\tau_2$.
 This happens when some state  has negative $H$,
i.e. when a tachyon first appears in the spectrum (apart from the usual bosonic string tachyon).
By examining the form of the Hamiltonian, one immediately sees that the first tachyon that appears as the temperature $T=(2\pi R_0)^{-1}$ is increased from zero
has $N_L=N_R=0$, $m_0=0$ and $w_0=\pm 1$. For such states, the critical $R_0$ is 
determined by
\eqn\hag{
H=0={R_0^2\over \a' }-2(a_L+a_R)\ ,
}
whereby
\eqn\hage{
T_H={1\over 2\pi R_0} = {1\over 2\pi \sqrt{ 2\a' (a_L+a_R)} } \ .
}

In the NSR formulation of type II superstring theory the calculation is 
similar. The tachyon appears in the NS-NS sector, where the
normal-ordering constants are 
$a_L=a_R=1/2$.
GSO projection does not remove this
tachyon state, because for odd winding number the GSO condition is reversed \aw\ 
(this is explicit in the one-loop expression for the free energy in sect. 2; the tachyon state with $w_0=0$ is projected out by GSO, but
not this thermal tachyon with $w_0=\pm 1$).

In order to reproduce this calculation in the Green-Schwarz formulation of the superstring 
(which is more suitable for the generalization
to membrane theory), we need to calculate
the normal ordering constant for the Euclidean theory
on $R^9\times S^1$. At zero temperature, the normal ordering constant
vanishes because of a cancellation between bosons and fermions.
In the thermal ensemble at finite temperature, fermions obey
antiperiodic boundary conditions under $X^0\to X^0+2\pi R_0$.
As a result, supersymmetry is broken and the vacuum energy will not vanish.
In type II superstring theory with 
 antiperiodic fermions, the number operators in the sector 
$w_0=\pm 1$ are given by
\eqn\anti{
N_L=\sum_{n=1}^\infty \big[\a_{-n}^i \a _n^i +(n-{1\over 2}) S_{-n}^a S^a_n \big]\ ,
\ \ \ N_R=\sum_{n=1}^\infty \big[\tilde \a_{-n}^i \tilde \a _n^i +(n-{1\over 2})
\tilde S_{-n}^a \tilde S^a_n\big]\ ,
}
$$
i=1,...,8\ ,\ \ \ \ \ \ a=1,...,8\ .
$$
Thus the normal ordering constant is as in the NS sector of the  NSR formulation,
i.e. $a_L=a_R= {1\over 2}$. In this way we reproduce the result for the Hagedorn temperature in the Green-Schwarz formulation.

The calculation of the normal ordering constants can be done by $\zeta $-function regularization. 
For the operators in \anti , one has
$$
a_L=a_R=-{1\over 2}(D-2) \big[ \ec_{\rm B} + \ec_{\rm F} \big]\ ,
$$
with 
$$
\ec_{\rm B} =\sum_{n=1}^ \infty n \ ,\ \ \ \ \ \ec_{\rm F}= - \sum_{n=0}^\infty (n+\ha )\ .
$$
Using the formulas
$$
\sum_{n=1}^ \infty {1\over n^\nu }=\zeta (\nu ) \ ,\ \ \ \  
\sum_{n=0}^\infty {1\over (n+ {1\over 2} )^\nu } =(2^\nu -1)\zeta (\nu )\ ,
$$
and $\zeta (-1) = -{1\over 12} $, we find
$$
\ec_{\rm B} =-{1\over 12} \ ,\ \ \ \ \ \ec_{\rm F}= -{1\over 24} \ ,
$$
so that for $D=10$ one has 
$a_L=a_R=\ha $.

\newsec{M-Theory at finite temperature}


We will describe M-theory at finite temperature as in sect.~2 by considering the 
 eleven-dimensional theory in euclidean target space with periodic time $X^0$, and periodic coordinate $X^{11}$.
Fermions are antiperiodic around $X^0$ and periodic around $X^{11}$.
Thus we are to consider  a toroidal compactification of Euclidean M-theory with $(-,+)$~ spin structure.  
Having the topology $R^9\times T^2$, membranes can 
wrap on a 2-torus $X^0,X^{11}$.

When viewed in eleven dimensions, the winding string that in sect. 3 led to a tachyon instability is a membrane  wrapped around $X^0,X^{11}$ with   winding number equal to $\pm 1$. Small oscillations of this membrane are effectively described by the $D=11$ supermembrane theory \bergsh . The observation that there is a tachyon instability at $T>T_{\rm cr}$ will be independent
of many details of the membrane Hamiltonian, depending  
only  on the net vacuum energy.
Although we will not determine the exact spectrum, being outside of the scope of this paper, it is interesting to note that,  because of the supersymmetry breaking boundary conditions, no flat direction  remains in the membrane Hamiltonian, so the exact supermembrane spectrum must be discrete (see discussion in sect. 4.2).

In the sector with zero winding, there can be other types of configurations which
give rise to low energy excitations, related to D0 brane configurations.
We emphasize, however,
that the aim here is not to provide a complete account of all
relevant excitations of M-theory at a given temperature 
and radius $R_{11}$, but rather to point out
the existence of a winding mode that becomes tachyonic at some critical temperature 
$T_{\rm cr}$ (which will depend on the radius of the eleventh dimension).
This does not exclude that there could be other instabilities.
In particular, in matrix model calculations it has been shown \sath\ that
at some sufficiently high temperature  there are D0 branes which
cluster at one point. This configuration might lead to a gravitational instability,
but estimating the temperature at which such configuration occurs does not appear to be 
simple \sath .

\subsec{Toroidal membranes}

Before considering the finite temperature case, it is convenient
to briefly review the light-cone Hamiltonian formalism
for membranes wrapped on a torus in Minkowski space, where $X^{10}$ and $X^{11}$ are compact.
Let $\s ,\rho \in [0,2\pi ) $ be the membrane world-volume coordinates. We can write
\eqn\qqq{
X^{10}(\s ,\rho )=w_0R_{10}\sigma + \tilde X^{10} (\s ,\rho )\ ,\ \ \  \ 
X^{11} (\s ,\rho )=\rr \rho  + \tilde X^{11} (\s ,\rho )\ ,
}
where $\tilde X^{10},\ \tilde X^{11}$ are single-valued functions of $\s $ and $\rho $.
The transverse coordinates $X^i(\s ,\rho )$, $i=1,2,...,8$ are all single-valued 
(we use the notation where the eleven bosonic coordinates are $\{ X^0, X^i,X^{10} , X^{11} \} $). They can be expanded in a complete basis of functions on the torus,
$$
X^i(\s,\rho )=\sqrt{\a '} \sum_{k,m} X^i_{(k,m)} e^{ik\s +im\rho }\ ,\ \ \ \  P^i(\s,\rho )={1\over (2\pi )^2 \sqrt{\a '}} \sum_{k,m} P^i_{(k,m)} e^{ik\s +im\rho }\ ,
$$
\eqn\tenn{
\a'=\big( 4\pi ^2 R_{11} T_{2}\big)^{-1}\ ,
}
where $T_{2}$ is the membrane tension ($[T_{2}]=cm^{-3}$). 
 The membrane light-cone Hamiltonian \refs{\bst,\dewitt }\ 
takes the form $H=H_0+H_{\rm int}$ , with \refs{\russo ,\constru } 
\def\ww{\omega_{km}}
\def\mn{ {(-k,-m)} }
\def\n{ {\bf n} }
$$
\a'  H_0= 8\pi^4 \a 'T_{2}^2 R_{10} ^2 R_{11}^2 w_0^2+  {1\ov 2} \sum _\n \big[ P_\n^i P^i_{-\n}
+\ww ^2 X^i_\n X^i_{-\n }\big]
$$
$$
\a'  H_{\rm int}= {1\ov 4g^2_A}\sum (\n_1 \times \n_2)(\n_3\times \n_4)
X_{\n_1}^i  X_{\n_2 }^j  X_{\n_3}^i   X_{\n_4}^j  
$$
$$
X^+={X^0+\td X^{11}\ov \sqrt{2} }=x^+ +\a' p^+\tau \ ,\ \ \ 
$$ 
$$
\n \equiv (k,m)\ ,\ \ \ \ \ \ 
\n \times\ \n '=km'-m k'\ ,\ 
$$
\eqn\tennn{
g^2_A\equiv {R_{11}^2\ov \a' }=4\pi^2R_{11}^3T_{2}\ ,\ \ \ \ \ \ 
\ww =\sqrt { k^2 + w_{0}^2 m^2 {R_{10}^2\ov R_{11} ^2}  }\ .
}
Here only the bosonic modes have been written explicitly (fermion modes will be included later). The constant $g_A$ represents the type IIA
string coupling. One can introduce
mode operators as follows:
\def\www {w _{ {(k,m)} } }
\eqn\modoz{
X^i_{(k,m)}={i\ov \sqrt {2} \www }\big[\a^i_{(k,m)}+\td \a^i_{(-k,-m)}\big]
\ ,\ \ \ \ P^i_{(k,m)}={1\ov \sqrt {2} }\big[\a^i_{(k,m)}-\td \a^i_{(-k,-m)}\big]\ ,
}
$$
\big( X_{(k,m)}^i\big) ^\dagger =X_{(-k,-m)}^i\ ,\ \ \ \ 
\big( P_{(k,m)}^i\big) ^\dagger =P_{(-k,-m)}^i\ ,\ \ \ \ 
\www \equiv  \epsilon (k ) \ \ww  \ ,
$$
where $\epsilon (k )$ is the sign function.
The canonical commutation relations imply
$$
\big[ X^i_{(k,m)} , P^j_{(k',m')} \big]= i\delta_{k+k'}\delta_{m+m'}\delta^{ij}\ ,
$$
\eqn\zzs{
[ \a _{ {(k,m)} }^i , \a^j_{(k',m')}]= \www \delta _{k+k'}\delta _{m+m'}\delta^{ij}\ ,\  
}
and similar relations  for the $\tilde \a _{ {(k,m)} } ^i$.

The  mass  operator is given by
\eqn\massa{
M^2=2p^+p^--(p^{i})^2-p_{10}^2=2 H_0+2 H_{\rm int}-(p^{i})^2-p_{10}^2\ .
}
The Hamiltonian is non-linear. There are two situations where one can
extract useful information from this Hamiltonian.
One is the limit of large $g_A$, with $R_{10}/R_{11}$ fixed,
in which the non-linear terms are multiplied by the  small number ${1\over g_A^2}$ and
 can be considered in perturbation theory. 
The other limit is $g_A\to 0$ at fixed $R_{10}/R_{11}$. This is related to the zero torus area limit 
of M-theory on $T^2$, which leads to
ten-dimensional type IIB string theory.
$H_{\rm int}$ is positive definite, and any state $|\Psi \rangle $ with 
$\langle \Psi | H_{\rm int} |\Psi \rangle \neq 0$ will have 
infinite mass in the zero area limit, where $g_A\to 0$
(with $T_{2}\to \infty $, so that $\a ' =(4\pi ^2 R_{11} T_{2})^{-1}$ and 
$R_{10}/R_{11} $ remain fixed).
The only states that survive  are
those states made of operators 
$\a _{ {n( p, q)} }^i,\ \tilde \a _{ {n( p, q)} } ^i$  
with the same value of $( p, q)$ \constru . They  satisfy
$
\langle \Psi | H_{\rm int} |\Psi \rangle = 0\ ,$
so that $H_{\rm int}$ drops out from
$\langle \Psi | M^2 |\Psi \rangle $. They describe the $(p,q)$ strings of type IIB superstring theory (the proposal that the $(p,q)$ string bound states of type IIB string theory originate from membranes was first made by Schwarz \schw ).

\def\n{ {\bf n} }

Let us now focus on the situation of large $g_A$.
To leading order  in perturbation theory in $1/g_A^2$, 
the interaction term can be dropped. 
The solution to the membrane equations of motion is given by
\eqn\xsol{
X^i (\s, \rho, \tau )= x^i +\a'  p^i \tau + 
i \sqrt{\textstyle {\a' \ov 2}} \sum_{\n \neq (0,0)}
 w_\n \inv \big[ \a _\n^i e^{ik\s +im\rho } 
+ \td \a _\n ^i e^{-ik\s -im\rho }\big] \ e^{i w_\n \tau } \ .
}
Let the momentum components 
in the directions $X^{10}$ and $X^{11}$ be given by 
$$
p_{10}={l _{10}\ov R_{10} }\ ,\ \ \ \ p_{11}={l _{11} \ov R_{11} }\ ,\
$$
where $l_{10},l_{11}$ are integers. The nine-dimensional mass operator takes the form $M^2={\cal H}$, with
\eqn\nima{
{\cal H}= {l^2_{10}\ov R_{10}^2} + {l^2_{11}\ov R_{11}^2} + {w_0^2 R_{10}^2\ov 
\a  ^{\prime 2}}\  + {1\ov \a' } {\bf H} \ ,\ \ \ \ 
}
\eqn\hhkk{
{\bf H}=  \sum _{k,m} \big( \a^i_{(-k,-m)} \a^i_{ (k,m)} + \td \a^i_{(-k,-m)} \td \a^i_{(k,m)}\big)\ .
}
The level-matching conditions are given by \refs{\duf ,\russo}
\eqn\cco{
N_\s^+ - N_\s^- = w_0 l_{10}\ ,\ \ \ \ \ N_\rho^+ - N_\rho ^- = l_{11} \ ,
}
where
$$
N^+_\s = \sum _{m=-\infty }^\infty \sum _{k=1}^\infty {k\ov \ww }
\a^i_\mn \a^i_{(k,m)} \ ,
\ \ \ N^-_\s = \sum _{m=-\infty }^\infty \sum _{k=1}^\infty {k\ov \ww }
\td \a^i_\mn \td \a^i_{(k,m)} \ ,
$$
$$
N^+_\rho=\sum _{m=1}^\infty \sum _{k=0}^\infty {m\ov \ww }
\big[ \a^i_\mn \a^i_{(k,m)} + \td \a^i_{(-k,m)} \td \a^i_{(k,-m)} \big]\ ,
$$
$$
N^-_\rho=\sum _{m=1}^\infty \sum _{k=0}^\infty {m\ov \ww }
\big[ \a^i_ {(-k,m)}\a^i_{(k,-m)} + \td \a^i_{(-k,-m)} \td \a^i_{(k,m)} \big]\ .
$$

To define the operator ${\bf H}$ in the quantum theory, we have to specify the normal ordering prescription. This will give rise to a vacuum energy. The annihilation operators are $\a _{ {(k,m)} }^i , \ \tilde \a _{ {(k,m)} }^i $ with $k>0$ for
all $m$, and $k=0$, $m>0$. Defining
\eqn\norm{
{\bf \hat H}= \sum _\n \big(: \a^i_{(-k,-m)} \a^i_{ (k,m)} : + :\td \a^i_{(-k,-m)} \td \a^i_{ (k,m) }:\big)\ ,\ \ 
}
where the normal ordering symbol ``::" means as usual taking the annihilation operators to the right, one finds the relation
\eqn\hhgg{
 {\bf H}=  {\bf \hat H}+ 2(D-3) \ec \ ,\ \ \ \  
}
$$
\ec = {1\over 2} \sum_{k,m} \om_{km}\ .
$$
This constant shift represents the purely bosonic contribution to the vacuum energy of the 2+1 dimensional field theory (discussed in \fuji ). If one chooses supersymmetry preserving boundary conditions for fermions, then the fermion and boson contributions to the vacuum energy cancel out \refs{\duf, \bergsh }. Being a consequence of the underlying supersymmetry, this result also holds when non-linear terms are included.

\subsec{ Vacuum energy for the finite temperature theory}

Let us now extend this to the supermembrane theory at finite temperature. The euclidean time coordinate $X^0$ plays
role of $X^{10}$. Fermions will
obey antiperiodic boundary conditions around $X^0$, and periodic boundary conditions around 
$X^{11}$.
We are interested in the sector $w_0=\pm 1$,  
where fermions are antiperiodic under $\s \to \s+2\pi $.
This implies that the frequencies $k$ in the Fourier expansions will be half integers, and the frequencies $m$ will be integers (since fermions are periodic under $\rho\to\rho+2\pi $). 
The Hamiltonian operator is (cf.~\anti )
\eqn\hhf{
{\cal H}= {l^2_{0}\ov R_{0}^2} + {l^2_{11}\ov R_{11}^2} + {R_{0}^2\ov 
\a  ^{\prime 2}}\  + {1\ov \a' } ({\bf \hat H} + 2(D-3) \ec ) \ ,\ \ \ \ 
}
where 
$$
{\bf \hat H}=  \sum _\n \big[ :\a^i_{-\n} \a^i_{\n} :+ :\td \a^i_{-\n} \td \a^i_{\n}: +
\om_{k+ {1\over 2} ,m} \big( :S^a_{-\n} S^a_{\n} : + :\td S^a_{-\n} \td S^a_{\n}:\big)\big]\ \  ,
$$
and 
\eqn\ffer{
\ec = \ec_{\rm B}+ \ec_{\rm F}= {1\over 2} \sum_{k,m} \big( \om_{km}- \om_{k+{1\over 2}, m} \big) \  ,
}
\eqn\fff{
\om_{km}=\bigg({k^2} +{m^2\over \gef ^2} \bigg)^{1\over 2}\ .
}

The sums in \ffer\ are divergent, but  they can be defined by analytic continuation. The procedure generalizes the
 usual zeta-function regularization used in sect. 3 for the superstring case, and it is equivalent to the functional relation $E_\nu (\Omega )=c E_{1-\nu }(\Omega )$ allowing the definition of Eisenstein series with $\nu <1/2 $ \terras .
We write
\eqn\com{
\eqalign{
\ec &=\lim _{\nu \to -{1\over 2} }\ {1\over 2} \sum_{k,m} \bigg( {1\over (\om _{km})^{2\nu } } - {1\over (\om _{k+{1\over 2},m})^{2\nu } }\bigg) \cr
&={\pi^\nu\over 2\Gamma (\nu ) }  \sum_{k,m}  \int_0^\infty {d\tau\over\tau } \tau^\nu \bigg( e^{-\pi \tau \om^2_{km}}-  e^{-\pi \tau (\om_{k+{1\over 2} ,m})^2}\bigg)
 \ .\ \ \ \ \cr }
}
Then, using the Poisson formula \pois\ 
one obtains
\eqn\mii{
\ec = 
 {\pi ^\nu \gef \over 2\Gamma (\nu ) } \sum_{w,w}\int _0^\infty {ds\over s} \ s ^{1-\nu } 
\big(1- (-1)^w \big) \exp \big[ -\pi s (w^2 +{w'}^2 \gef ^2 )\big]\ ,
}
where we have made the change of integration variable, $s=1/\tau $.
Hence 
\eqn\sii{
\ec =
{\gef  \pi ^{2\nu -1}\Gamma (1-\nu ) \over 2\Gamma (\nu )   }
\sum_{w,w'} \big( 1- (-1)^w \big)  {1\over (w^2 +{w'}^2 \gef ^2 )^{1- \nu }} \ .
}
Setting now $\nu =-1/2 $ we get
\eqn\siif{
\ec =
- {\gef  \over 8 \pi ^2   }
\sum_{(w,w') \neq (0,0)}    \big( 1- (-1)^w \big)     (w^2 +{w'}^2 \gef ^2 )^{-{3\over 2}} \ .
}
Thus we have (see eqs. (A.1), (B.1))
\eqn\eee{
\ec = - {1 \over 4 \pi ^2 \sqrt{\gef }   }\ \zeta(3) \big( E_{3\over 2}  (\gef )- F_{3\over 2}  (\gef )\big) \ .
}


Using the expansions (A.5), (A.6), (B.9), (B.10) the  vacuum energy takes the form
\eqn\aah{
\ec = -{1\over 8} +O(e^{-2\pi/\gef }) \ ,\ \ \ \ \ \gef\ll 1\ ,
}
and
\eqn\aahh{
\ec = -{7\over 16\pi^2 }\zeta(3)\gef -{1\over 12\gef } +O(e^{-2\pi \gef })\ ,
\ \ \ \ \gef \gg 1\ .
}
The explicit analytic form for the exponentially small terms can be read from the formulas in the appendices.

Notably, eq.~\aah\ implies that at $\gef \ll 1$ (i.e. small type IIA coupling $g_A$ or sufficiently low temperatures), the vacuum energy is identical to that of type II superstring theory, 
i.e. $2(D-2)\ec =-2(D-2)({1\over 12}+ {1\over 24})$, due to a cancellation
of the term proportional to $\zeta (3)/\gef ^{3/2}$ in the expansions (A.5), (B.9).

A question is why the vacuum energy 
gives the correct result in the weak coupling limit $g_A\ll 1$.
 In the derivation of the vacuum energy, 
we have used the assumption that $g_A\gg 1$ to neglect the contribution of the non-linear terms. The fact that the correct result emerges at weak coupling indicates that a possible extra contribution coming from the non-linear terms in the Hamiltonian may cancel out between fermion and boson contributions. 

Another interesting point is the issue of flat directions in the membrane Hamiltonian for the wrapped membrane. Consider first the case of supersymmetric boundary conditions. In the strict limit $g_A\to \infty $, one has a Hamiltonian which is a sum of  harmonic oscillators, so there is no flat direction and the spectrum is discrete. For any finite $g_A\gg 1$, states representing small oscillations should be almost stable, since they only see the harmonic potential. However, if flat directions are present, they may decay by tunnel effect (see also discussion in \plefka ). This effect should be exponentially small for large $g_A$.
Now, in the present case of non-supersymmetric boundary conditions,
possible flat directions will be removed by the same effect flat directions are removed in the bosonic theory (described in \dln ).
The motion is confined to some finite region, and  the exact spectrum
of the supermembrane must be discrete. For large values of $g_A$,
most of the states are confined to the harmonic region of the potential, so this effect should not play a significant role.

\subsec{Critical temperature in M Theory}

As in the string theory case, there will be a tachyonic instability when ${\cal H}<0$ for some state (see \hhf ). 
The  first state that solves ${\cal H}=0$ is a state 
 with $l_0=l_{11}=0$, $w_0=\pm 1$, which is annihilated by all annihilation operators $\a^i_{\n},\ \td \a^i_{\n} $ 
(this is nothing but the ``uplift" of the winding tachyon of string theory to eleven dimensions).
{}From eq.~\hhf , we thus find that the critical temperature is determined by the solution of the equation
\eqn\aaa{
{\cal H}=0= {R_0^2\over \a'} + 2(D-3)\ec \ .
}
Using eq.~\eee\ and setting $D=11$, this becomes 
\eqn\crtt{
  {1\over T^2_{\rm cr}}=   \a ' {16 \over   \sqrt{\gef }   }\ \zeta(3) \big( E_{3\over 2}  (\gef )- F_{3\over 2}  (\gef )\big) \ .
}
Here $\gef = 2\pi \sqrt{\a '} T_{\rm cr} g_A$, so this is a transcendental equation for $T_{\rm cr}$.
It  can be solved analytically in two regimes,  $\gef \ll 1 $ and $\gef \gg 1 $, using the expansions \aah\  and \aahh .
At weak coupling $\gef \ll 1$, the equation \aaa\ becomes
\eqn\ggt{
{R_0^2\over \a'} = - 16\ \left( -{1\over 8} + O(e^{-{2\pi\over \gef } }) \right) \ , 
}
i.e.
\eqn\tcrrr{
T_{\rm cr}={1\over 2\pi \sqrt{2\a' }} \ .
}
This coincides with the Hagedorn temperature.  This is a consequence of the observation of the previous subsection  that the vacuum energy reduces to the weak coupling superstring value at $\gef \ll 1$.
Since the critical temperature in this regime is of order $1/\sqrt{\a '}$, the condition $\gef \ll 1$ implies $g_A \ll 1$.

It is easy to see that in a regime $g_A\gg 1$, the coupling $\gef $ will be large at the critical temperature $T=T_{\rm cr}$. This means that, in order to determine the critical temperature, we have to use \aahh . 
Using \aahh\ and keeping only the leading term, the condition \crtt\  determining the critical temperature  becomes
\eqn\ttcr{
{1\over T_{\rm cr}^2}\cong \a ' 28\zeta (3)\gef = 56 \pi {\a '}^{3/2}
\zeta (3) g_A \ T_{\rm cr}  \ .
}
Thus  the critical temperature at strong coupling is
\eqn\ttaa{
T _{\rm cr} \cong {1\over a \sqrt{\a' } (2\pi g_A)^{1\over 3} } = 
{1\over a} \big(2\pi T_2 \big)^{1/3}
\ ,\ \ \ \ g_A\gg 1\ ,
}
$$
a= \big[ 
28 \zeta (3)   \big]^{1\over 3}\cong 3.23 \ ,
$$
where $T_2$ is the membrane tension (we have used eqs.~\tenn, \tennn ).
In terms of the eleven-dimensional Planck length $l_P$, 
$g_A^2=2\pi {R^3_{11}\over l_P^{3}}$, so  
\eqn\tee{
T_{\rm cr}={1\over a\  l_P}\cong 0.31\ l_P^{-1}\ \ .
}
Here $l_P$ is normalized so that the gravitational coupling is 
$\kappa_{11}^2=16\pi^5 l_P^9 $.
As a check, note that $\gef (T_{\rm cr}) =
2\pi\sqrt{\a' } T_{\rm cr} g_A \sim g_A^{2/3}\gg 1$, which is consistent with 
approximating the Eisenstein functions by eq.~\aahh . The same result \ttaa\ is obtained by solving \crtt\ numerically at $g_A\gg 1$.
Thus the critical temperature for the type IIA superstring decreases at strong coupling. In terms of the eleven-dimensional
Planck length, at large radius, it approaches a constant value, $T_{\rm cr}\cong 0.31\  l_P^{-1}$.

By studying an expression for the free energy computed in a semiclassical approximation,
in ref.~\odin\ a ``regularized" Hagedorn temperature was proposed for
the $D=11 $ theory, which becomes infinity as the cutoff 
is sent to zero. The numerical
coefficient contains a similar factor $7 \zeta (3)$ as in \ttcr. 
The appearance of this factor in \odin\ is also related to the vacuum energy of the 
world-volume theory in $D=11$.

\smallskip

The critical temperature can be obtained for all values of the coupling by solving eq.~\crtt\ numerically.
Fig.~1 is a plot of the critical temperature as a function of $\gef $, and fig.~2 is a plot of $T_{\rm cr}$ as a function of $g_A$.
At small couplings, the plots have the same behavior, since $T_{\rm cr}$ is approximately constant. At strong coupling, $T_{\rm cr}$ goes to zero as $1/\gef^{1/2} $ in fig.~1, and as $1/g_A^{1/3}$ in fig.~2.

\ifig\fone{Critical temperature as a function of $\gef $ (with $\a'=1 $). The value at $\gef=0 $ is $T_H=1/(2\pi\sqrt{2\a '})$. }
{\epsfxsize=5.0cm \epsfysize=5.0cm \epsfbox{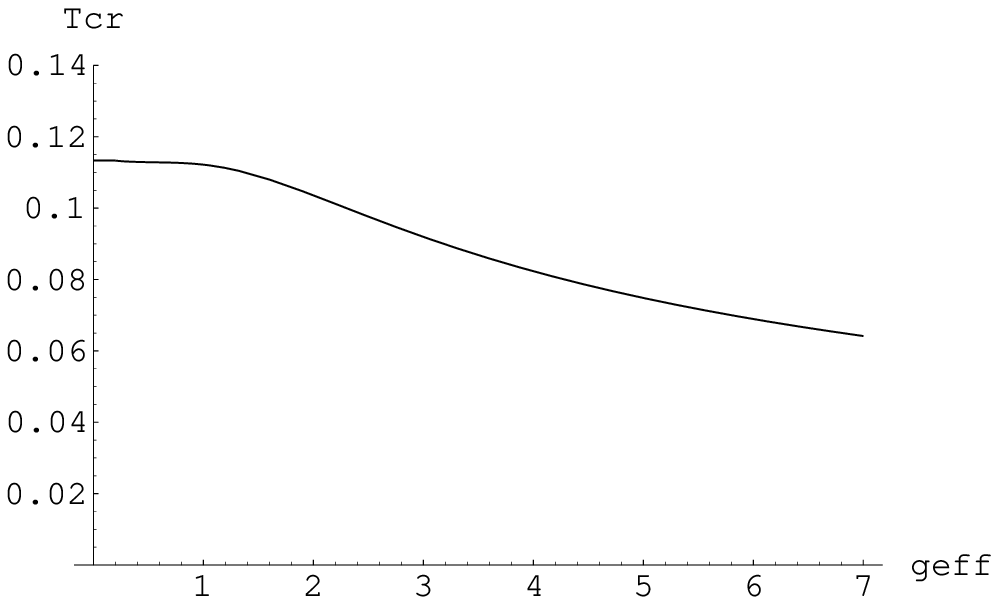}}

\ifig\fone{Critical temperature as a function of $g_A$.}
{\epsfxsize=5.0cm \epsfysize=5.0cm \epsfbox{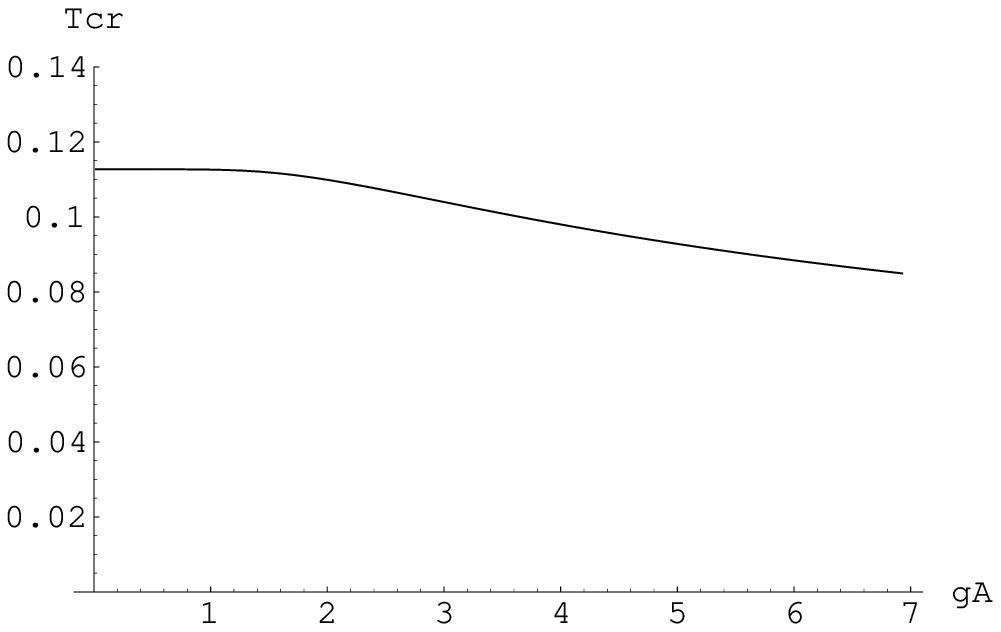}}

The natural mass scale in eleven dimensions is $l_P^{-1}$. This means that the energies of elementary excitations of the system at large $g_A$ must be proportional to
$l_P^{-1}$, not $1/\sqrt{ \a '}$.
So temperature at $g_A\gg 1$ is more properly measured in units of $l_P^{-1}$.
An analogous situation happens in type IIB superstring theory.
 The Hagedorn temperature at $g_B\gg 1$ can be
found by S-duality.
The strong coupling limit of type IIB theory is known to be the same theory, where fundamental strings are replaced by D-strings, $g_B\to 1/g_B$ and  $\a' $ by $\a'_D =g_B\a' $, so that the Hagedorn termperature is 
$$
T_{H}= {1\over 2\pi \sqrt{2\a'_D }}= {1\over 2\pi \sqrt{2\a' g_B}}\ .
$$
The Hagedorn temperature  goes to zero for large $g_B$ at fixed $\a' $, 
but it should be measured with respect to
the D-string tension (since elementary excitations have energies of order $1/\sqrt{\a'_D}$), in which case it is a constant independent of the coupling. In principle, it seems possible to generalize the present method to determine
 a critical temperature in type IIB superstring theory at some intermediate coupling $g_B$. In order to connect
M-theory at finite temperature with type IIB theory one needs to study M-theory on an euclidean 3-torus. Now membranes can wrap in different ways.


\subsec{Duality connections}

The Hagedorn temperature in string theory was first
understood as a consequence of the exponential growth of the asymptotic level density with the mass, $\rho (m) \sim e^{ {\rm const.} m}$.
The existence of a finite critical temperature
 at $g_A\gg 1$ can be explained
if there is a string-theoretic description of M-theory in this limit.
The tension $1/\tilde \a '$ of such string can be read (up to a numerical constant) from the critical temperature.
We have the formula:
\eqn\kko{
T_{\rm cr} ^2 = {1\over 28\zeta (3)\a' \gef }={ {\rm const.}\over \tilde\a'}\ .
}
How can a string-theory description arise at large $R_{11}$~? 
In this limit, near the critical temperature we have  $R_{0}\ll R_{11}$.
Therefore, the relevant low energy degrees of freedom of the system  are more appropriately described by  making dimensional reduction along the Euclidean time direction $X^0$, 
with $X^{11}$ now playing the role of a compact spatial dimension  of the resulting ten-dimensional theory. Because of the antiperiodic boundary conditions around $X^0$, 
the resulting ten dimensional string theory is a non-supersymmetric string theory. According to \berg , 
a compactification of M-theory on a circle where  fermions obey antiperiodic boundary conditions gives type 0A string theory  
(the relation between finite temperature type IIA theory  and type 0A theory was 
noted already in \aw  ).
Therefore the strong coupling limit of type IIA superstring theory at finite  temperature $T$ would 
be described by euclidean type 0A string theory, where the string coupling is 
$\tilde g_A^2= 2\pi {R_0^3\over l_P^3}=(4\pi^2 T^3l_P^3)^{-1}$, and the string tension
is obtained from the usual formula 
$\tilde \a'= {l_P^3\over 2\pi R_0}=l_P^3 T=\a' \gef $.
This  agrees with the identification in \kko . 
In the strict limit $R_{11}\to \infty $, this duality 
implies that {\it uncompactified} M-theory at  temperature $T$ 
is described by a ten-dimensional euclidean string theory.

Reproducing
the numerical coefficient in \kko \ using string theory techniques may not be simple, because the type 0A theory is strongly coupled below the critical temperature, i.e. 
$\tilde g_A>O(1) $ for $T<T_{\rm cr}$.
However, eq. \kko\ predicts that the type 0A tachyon must disappear
at a coupling
$$
\tilde g_A^2 > \tilde g^2_{A{\rm cr}}={7 \zeta (3) \over \pi^2} \cong 0.85\ .
$$
The precise numerical value may be subject to corrections,
for reasons explained in sect.~1. 
This agrees with the suggestion of \berg , that the type 0A tachyon should become massive at strong coupling. Conversely,  the existence of
a critical coupling $\tilde g_{A\rm cr} $ in type 0A string theory 
at which the type 0A tachyon becomes massless 
implies  the existence of the critical temperature $T=O(l_P^{-1})$ found in this paper
in uncompactified M-theory at finite temperature.
In terms of the critical coupling, the critical temperature is 
$T_{\rm cr}= (2\pi g_{A\rm cr})^{-2/3}\ l_P^{-1}$.
It would be 
interesting to investigate further  consequences of this connection in more detail.

\bs\bs

\noindent{\bf Acknowledgements}
\medskip
This work is supported by Universidad de Buenos Aires 
and Conicet.


\appendix {A}{Non-holomorphic Eisenstein series}

The non-holomorphic Eisenstein series is defined by \terras 
\eqn\ess{
2 \zeta (2r ) E_r(\Omega )=\sum_{(k,m)\neq (0,0)} {\Omega_2^r\ov |k+m\Omega |^{2r} }\ ,
\ \ \ \ \ r>{1\over 2}\ ,
}
where $\Omega =\Omega_1+i\Omega_2 $ is a complex parameter describing the upper half 
complex plane.
The Eisenstein series $E_r(\Omega )$ is invariant under $SL(2,Z)$ 
transformations of $\Omega $,
\eqn\oop{
\Omega \to {a\Omega +b\over c\Omega+d}\ ,\ \ \ ad-bc=1\ ,\ \ \ a,b,c,d\ \in \ Z\ .
}
At large $\Omega _2 $, one has the expansion
\eqn\bee{
E_r(\Omega )=\Omega_2^r+\gamma_r \Omega_2^{1-r}+
{4\Omega_2^{1/2}\pi^r\ov\zeta(2r)\Gamma(r)}
\sum_{n,w=1}^\infty \big({w\ov n}\big)^{r-1/2}\cos(2\pi  wn\Omega_1)
K_{r-1/2}(2\pi w n\Omega_2 )\ ,
}
$$
\gamma_r={\sqrt{\pi }\ \Gamma(r-1/2)\ 
\zeta(2r-1)\ov \Gamma(r)\ \zeta(2 r) }\ .
$$
The derivation is as in the analogous case given in appendix B.
Using the asymptotic expansion for the Bessel function $K_{r-1/2}$,
\eqn\besk{
K_{r-1/2}(2\pi w n\Omega_2 )={1\ov \sqrt{4wn\Omega_2 } }e^{-2\pi w n\Omega_2}\sum_{m=0}^\infty {1\ov (4\pi wn \Omega_2)^m }{\Gamma(r+m)\ov \Gamma(r-m)m! }\ ,
}
we see that the terms in \bee\ involving the Bessel function will be
exponentially suppressed.
In the present case with $\Omega_2= {1\over \gef }$, such exponentially suppressed
terms
represent non-perturbative contributions originating from $D0$ branes, whose
coefficient is therefore exactly determined by the above expansion of the Bessel function.
In a strong coupling expansion --~obtained 
by the modular transformation $\Omega \to -\Omega ^{-1}$~-- the exponentially suppressed terms are instead of the form $e^{-2\pi\gef }$.

{}From equation \bee , we obtain the following expressions for the expansions
of Eisenstein series appearing in sects. 2 and 4:
\eqn\eet{
\zeta (3) E_{3\over 2}(\gef ) ={\zeta (3) \over \gef ^{3\over 2} } + {\pi^2\over 3}\ 
\gef ^{1\over 2} +{8\pi \over \sqrt{\gef  } }\sum_{n,w=1}^\infty   {w\over  n} 
K_1(2\pi {wn \over\gef} )\ , \ \ {\rm for}\ \ \gef\ll 1\ ,
}
\eqn\eett{
\zeta (3) E_{3\over 2}(\gef ) ={\zeta (3) \gef ^{3\over 2} } + {\pi^2\over 3}\ 
{1\over \gef ^{1\over 2} } +8\pi  \sqrt{\gef  } \sum_{n,w=1}^\infty   {w\over  n} 
K_1(2\pi wn \gef )\ ,\  \ {\rm for}\ \ \gef\gg 1\ ,
}
and
\eqn\deb{
\eqalign{
\zeta (11) E_{11\over 2}(\gef )&=\zeta (11) \gef^{-11/2} +{256\zeta (10) \over 315}
 \gef^{9/2} \cr
& + {4\pi ^{11/2} \over 
\Gamma({11\over 2})\sqrt{\gef  } }\sum_{n,w=1}^\infty   \big( {w\over  n}\big) ^5 
K_5(2\pi {wn \over\gef} )  \ ,\ \ \ \ \gef\ll 1\ , \cr }
}
\eqn\fue{
\eqalign{
\zeta (11) E_{11\over 2}(\gef )&=\zeta (11) \gef^{11/2} +{256\zeta (10) \over 315}
 \gef^{-9/2} \cr
&+ {4\pi ^{11/2} \sqrt{\gef  } \over \Gamma({11\over 2}) }\sum_{n,w=1}^\infty   \big({w\over  n}\big) ^5 
K_5(2\pi wn \gef )\ , \ \ \ \ \gef\gg 1\ .\cr }
}

\appendix {B}{Generalized Eisenstein series for fermion contributions}

In the calculations performed in the main text, fermion contributions 
(either to the free energy or to the vacuum energy) led to Eisenstein-type series 
of the form
\eqn\esss{
2 \zeta (2r ) F_r(\Omega )\equiv
\sum_{(k,m)\neq (0,0)} (-1)^m {\Omega_2^r\ov |k+m\Omega |^{2r} }\ ,\ \ \ r> \ha \ .
}
Here we will derive
some basic properties that we need, such as the weak and strong coupling expansions.

Note that $F_r(\Omega )$ is {\it not} $SL(2,Z)$ invariant,
so the weak and strong coupling expansions will be different.
In particular, the modular transformation $\Omega \to -1/\Omega $ gives
\eqn\essd{
2 \zeta (2r ) F_r(-1/\Omega )\equiv
\sum_{(k,m)\neq (0,0)} (-1)^k {\Omega_2^r\ov |k+m\Omega |^{2r} }\ .
}

Let us first derive an expansion of \esss\ for $\Omega_2\gg 1$.
We will make use of the formulas:
$$
\sum_{n=1}^\infty {1\over n^s}=\zeta( s)\ ,\ \ \ \ 
\sum_{n=1}^\infty {(-1)^n\over n^s}=-\zeta( s) (1-2^{1-s})\ .
$$
Separating the $m=0$ term in eq.~\esss , we get
\eqn\solo{
 \zeta (2r ) F_r(\Omega )=\zeta(2r) \Omega_2^r
+ \Omega_2^r
 \sum_{k}\sum _{m = 1}^\infty (-1)^m {\pi ^r\over \Gamma(r)}
\int_0^\infty {dx\over x} x^r e^{-\pi x |k+m\Omega |^2}\ .
}
We now use the Poisson resummation formula,
\eqn\popo{
\sum_k f(k)=\sum_{k'}\int_{-\infty}^\infty d\mu\ f(\mu )\ e^{2\pi i\mu k'} \ .
}
 We get
\eqn\sola{
\zeta (2r ) F_r(\Omega )=\zeta(2r) \Omega_2^r + {\Omega_2^r\pi ^r\over \Gamma(r)}
 \sum_{k'}\sum _{m = 1}^\infty (-1)^m e^{2\pi i k' m \Omega_1} 
\int_0^\infty {dx\over x} x^{r- {1\over 2}}
e^{-\pi x m^2\Omega_2^2   - { \pi {k'}^2\over x}}
}
Separating the term $k'=0$, and performing the integrations, we finally obtain
\eqn\prr{
\eqalign{
\zeta (2r ) F_r(\Omega )&= \zeta(2r) \Omega_2^r
+\beta_r \Omega_2^{1-r}
\cr
&+{4\Omega_2^{1/2}\pi^r\ov \Gamma(r)}
\sum_{k=1}^\infty\sum_{m=1}^\infty (-1)^m \big( {k\ov m}\big)^{r-1/2}\cos(2\pi  km\Omega_1)K_{r-1/2}(2\pi km\Omega_2 )\ ,
\cr }
}
$$
\beta_r \equiv -{\sqrt{\pi } \Gamma(r-1/2)
\ov \Gamma(r) }\zeta(2r-1)\ \big( 1-2^{2-2r} \big) \ .
$$
For $\Omega =i/\gef $, eq.~\prr\ is an expansion which is applicable in the regime $\gef \ll 1$. An expansion for the opposite regime, $\gef \gg 1$, can be obtained by proceeding in a similar way, but
separating the $k=0$ term in eq.~\esss.
Define $\tilde \Omega=-1/\Omega $, so that $\tilde \Omega_2={\Omega_2 \ov |\Omega|^2}\ ,\ \ 
\tilde \Omega_1=-{\Omega_1 \ov |\Omega|^2 }$. 
We get
\eqn\ssee{
 \zeta (2r ) F_r(\Omega )=-\zeta(2r)(1-2^{1-2r}) \tilde \Omega_2^r
+ \tilde \Omega_2^r
 \sum_{m}\sum _{k = 1}^\infty (-1)^m {\pi ^r\over \Gamma(r)}
\int_0^\infty {dx\over x} x^r e^{-\pi x |m+k\tilde \Omega |^2}\ .
}
Now we make Poisson resummation in $m$, and then perform the integration over $x$.
We obtain
\eqn\ssdd{
\eqalign{
 \zeta (2r ) F_r(\Omega ) &= -\zeta(2r)(1-2^{1-2r}) \tilde \Omega_2^r
\cr
&+  {2\tilde \Omega_2^{1/2}\pi^r\ov \Gamma(r)}
\sum_m  \sum_ {k=1}^\infty \big({ |m+ \ha | \ov k}\big)^{r-{1\over 2}} 
e^{2\pi i k (m+{1\over 2})\tilde \Omega_1 }
K_{r-{1\over 2}} (2\pi k |m+ \ha |\tilde \Omega_2 )\ .\cr}
}
This expansion is applicable for large $\tilde \Omega_2 $. In our case, 
we have $\Omega =i/\gef $ and $\tilde \Omega = i \gef $, so eq.~\ssdd\ gives an expansion for  $\gef \gg 1$.
Note that there is only one power-like term, and the remaining terms are exponentially suppressed at large $\tilde\Omega_2 $.

Summarizing, we obtain for $r=3/2$ and $r=11/2$ the following expansions (cf.~\eet -- \fue )
\eqn\fes{
\zeta(3) F_{3\over 2}(\gef ) ={\zeta(3) \over \gef ^{3\over 2} } -{\pi^2\over 6}\gef^{1\over 2}
+{8\pi \over \sqrt{\gef  } }\sum_{k,m=1}^\infty (-1)^m  {k\over  m} 
K_1(2\pi {km \over\gef} )\ ,\ \ \ \ \gef\ll 1\ ,
}
\eqn\fess{
\zeta(3) F_{3\over 2}(\gef ) =-{3\over 4 } \zeta(3) \gef^{3\over 2}
+{4\pi \sqrt{\gef  } }\sum_{m}\sum_{k=1}^\infty {|m+ \ha |\over  k} 
K_1(2\pi \gef k|m+ \ha | )\ ,\ \  \gef\gg 1 ,
}
and
\eqn\qqo{
\eqalign{
\zeta(11) F_{11\over 2}(\gef ) &= {\zeta(11) \over \gef ^{11/2} }- 
 {256\over 315 }(1-2^{-9})\zeta (10) \gef^{9/2}
\cr
&+\ {4\pi ^{11/2} \over \Gamma({11\over 2}) \sqrt{\gef  } }\sum_{k,m=1}^\infty (-1)^m \big( {k\over  m} \big)^5
K_5(2\pi {km \over\gef} )\ ,\ \ \ \ \gef\ll 1\ , \cr }
}
\eqn\qqoo{
\eqalign{
\zeta(11) F_{11\over 2}(\gef ) &=  -(1-2^{-10})\zeta(11) \gef^{11/2}  
\cr
&+\ {2\pi^{11/2} \sqrt{\gef }\over \Gamma({11\over 2})  }\sum_{m}\sum_{k=1}^\infty 
{|m+{1\over 2}|^5 \over  k^5 } 
K_5(2\pi \gef k|m+ \ha | )\ ,\ \ \ \ \gef\gg 1\ .\cr }
}

\listrefs

\vfill\eject\end